\documentclass{article}

\usepackage{arxiv}

\usepackage[utf8]{inputenc} 
\usepackage[T1]{fontenc}    
\usepackage{hyperref}       
\usepackage{url}            
\usepackage{booktabs}       
\usepackage{amsfonts}       
\usepackage{nicefrac}       
\usepackage{microtype}      
\usepackage{graphicx}
\usepackage{natbib}
\usepackage{doi}

\usepackage[acronym,toc,shortcuts]{glossaries}
\makeglossaries
\newacronym{ics}{ICS}{Industrial Control Systems}
\newacronym{ot}{OT}{Operational Technology}
\newacronym{ctt}{CTT}{Cyber Table Top}
\newacronym{hvac}{HVAC}{heating, ventilation, and air conditioning}
\newacronym{cve}{CVE}{Common Vulnerabilities \& Exposures}
\newacronym{cwe}{CWE}{Common Weakness Enumerations}
\newacronym{capec}{CAPEC}{Common Attack Pattern Enumeration and Classification}
\newacronym{nist}{NIST}{National Institute of Standards and Technologies}
\newacronym{sp}{SP}{Special Publication}
\newacronym{sparta}{SPARTA}{Space Attack Research and Tactic Analysis}
\newacronym{mbse}{MBSE}{Model-Based System Engineering}
\newacronym{sbom}{SBOM}{Software Bill of Materials}
\newacronym{nvd}{NVD}{National Vulnerability Database}
\newacronym{marl}{MARL}{Multi-Agent Reinforcement Learning}
\newacronym{cia}{CIA}{Confidentiality, Integrity, and Availability}
\newacronym{rl}{RL}{Reinforcement Learning}
\newacronym{coa}{COA}{course of action}
\newacronym{xai}{XAI}{Explainable AI}
\newacronym{mdo}{MDO}{Multi-Domain Operations}
\newacronym{cvss}{CVSS}{Common Vulnerability Scoring System}
\newacronym{fived}{5D}{Deny, Degrade, Disrupt, Deceive, and Destroy}
\newacronym{ttp}{TTP}{tactics, techniques and procedures}
\newacronym{apt}{APT}{Advanced Persistent Threat}
\newacronym{cei}{\ensuremath{CE[I]}}{expected cumulative impact}
\newacronym{cert}{CERT}{Computer Emergency Response Team}
\newacronym{cin}{\ensuremath{C_{IN}}}{expected cumulative inbound impact}
\newacronym{cnode}{\ensuremath{C_{NODE}}}{expected nodewise cumulative impact}
\newacronym{cout}{\ensuremath{C_{OUT}}}{expected cumulative outbound impact}
\newacronym{cpt}{CPT}{Cyber Protection Team}
\newacronym{hci}{HCI}{Host Controller Interface}
\newacronym{msel}{MSEL}{major sequence of events list}
\newacronym{plc}{PLC}{Programmable Logic Controller}
\newacronym{jme}{SME}{Subject Matter Expert}
\newacronym{marlfw}{MARL Framework}{proprietary multi-agent reinforcement learning framework}
\newacronym{kgp}{KG Platform}{proprietary knowledge graph platform}
\newacronym{mlss}{MLSS}{proprietary machine learning based scoring system}

\usepackage{subfig}

\usepackage{hhline}
\usepackage{multirow}
\usepackage{makecell}

\usepackage[normalem]{ulem}
\usepackage{xcolor}

\newcommand{\crow}{Crow\texttrademark{}}
\newcommand{\vortex}{Vortex\texttrademark{}}

\graphicspath{{figures}}

\title{AI-Based Vulnerability Assessment Capability and Cyber Attack Graph Analysis}

\author{
	Joni Herttuainen \\
	Aalto University School of Science \\
	P.O. Box 11000, 00076 Aalto, Finland \\
	\texttt{joni.herttuainen@aalto.fi} \\
	\And
	Kirsi Hellsten \\
	Aalto University School of Science \\
	P.O. Box 11000, 00076 Aalto, Finland \\
	\texttt{kirsi.hellsten@aalto.fi} \\
	\And
	Vesa Kuikka \\
	Aalto University School of Science \\
	P.O. Box 11000, 00076 Aalto, Finland \\
	\texttt{vesa.kuikka@aalto.fi} \\
	\And
	Ambrose Kam \\
	Lockheed Martin \\
	\texttt{ambrose.kam@lmco.com} \\
	\And
	Arlanda Johnson \\
	Lockheed Martin \\
	\texttt{arlanda.e.johnson.jr@lmco.com} \\
	\And
	David Welsh \\
	Lockheed Martin \\
	\texttt{david.welsh@lmco.com} \\
	\And
	Kimmo K. Kaski \\
	Aalto University School of Science \\
	P.O. Box 11000, 00076 Aalto, Finland \\
	\texttt{kimmo.kaski@aalto.fi} \\
}

\renewcommand{\shorttitle}{AI-Based Vulnerability Assessment \& Attack Graphs}

\hypersetup{
pdftitle={AI-Based Vulnerability Assessment Capability and Cyber Attack Graph Analysis},
pdfsubject={cs.CR, cs.AI},
pdfauthor={Joni Herttuainen, Kirsi Hellsten, Vesa Kuikka, Ambrose Kam, Arlanda Johnson, David Welsh, Kimmo K. Kaski},
pdfkeywords={Attack Graphs, Multi-Agent Reinforcement Learning, Knowledge Graph, Cyber Risk Assessment, Vulnerability Assessment, Critical Infrastructure},
}

\begin{document}
\maketitle

\begin{abstract}

    Cyber threats targeting mission-critical infrastructure are becoming more sophisticated while the barrier to launching attacks continues to fall. Traditional point solutions like antivirus and firewalls are reactive and fail to address the combinatorial complexity of modern attack surfaces. This paper presents an  investigation combining two complementary methodologies: Lockheed Martin's \vortex/\crow{} framework, which applies multi-agent reinforcement learning (MARL) over industry-standard cyber knowledge graph to identify and prioritize attack vectors and TTPs (tactics, techniques, and procedures); and Aalto's probabilistic attack graph model that combines network topology and its vulnerabilities to compute system-level risk metrics. The 2015 Ukraine Power Grid cyberattack serves as a well-documented validation scenario. Applied independently to the same operational technology (OT) network topology, both methodologies converge on the same attack vectors and exploit sequences as those documented in the incident record, thus providing mutual cross-validation. Attack graph analyses using node-level elimination experiments identify industrial control systems (ICS) as the most critical enablers of attack propagation, representing high-priority targets for defensive hardening. Comparison of CVSS (v2.0) and IronMiner vulnerability scoring yields in general consistent results, with IronMiner providing more actionable differentiation at network periphery nodes. The layered methodology of baseline assessment and node-level elimination proves to be scalable to large enterprise networks, thus offering defenders a structured, AI-enabled path to prioritize mitigation under realistic time and resource constraints.

\end{abstract}


\section{Introduction}
\label{sec:intro}

Cybersecurity has become one of the most consequential challenges of modern societies, touching nearly every network domain, from personal devices and financial systems to industrial control systems (ICS), operational technology (OT) and critical infrastructure such as power generation, water treatment and transportation. Unlike conventional physical threats, cyberattacks require minimal resources to initiate as malware and ready-to-use exploitation tools are freely available, and recent conflicts in Ukraine and Gaza have shown that cyber operations can be used at scale as strategic instruments, disrupting civilian infrastructure with cascading societal and economic consequences. Ransomware campaigns against utilities, hospitals, and transportation networks further confirm that cyber threats extend far beyond espionage or financial motives.

Conventional defenses are poorly matched to this landscape. Traditional cybersecurity is largely a collection of reactive point solutions, e.g. antivirus, firewalls, and intrusion detection, that are narrow in scope, and even sound cyber-hygiene practices can fail against sophisticated, multi-stage attacks. For an enterprise OT network that spans hundreds or thousands of nodes, each exposing many exploitable services, an exhaustive manual assessment is impractical. Attack surface analysis, Red Team exercises, and Cyber Table Top (CTT) engagements yield valuable insight but are labor-intensive, hard to scale, and inherently incomplete given the combinatorial space of possible attack paths.

This motivates a shift toward proactive AI-enabled analysis that operates at the system-of-systems level, integrating vulnerability data, network topology, adversary behavior, and probabilistic risk metrics to anticipate the most impactful exploitation sequences and prioritize mitigations before an incident occurs. This paper reports a joint study by Lockheed Martin (LM) and Aalto University that pairs two independently developed but complementary methodologies toward this goal. LM's \vortex{} platform builds a cyber knowledge graph by ingesting system configuration data together with recognized vulnerability databases (CVE~\cite{cve}, CWE~\cite{cwe}, CAPEC~\cite{capec}) and cyber frameworks (NIST SP 800-53B~\cite{nist80053}, MITRE ATT\&CK~\cite{attck}, D3FEND~\cite{d3fend}, ATLAS~\cite{atlas}, FiGHT~\cite{fight}), upon which LM's \crow{} system trains multi-agent reinforcement learning (MARL) agents to discover optimal attack sequences aligned with specified mission objectives. This is complemented by Aalto University's probabilistic attack graph framework that models holistically network topology and exploitation of its vulnerabilities as a directed graph, enabling computation of path- and node-level risk metrics such as expected cumulative impact and inbound and outbound propagation risks. To validate and cross-compare these approaches, both were applied independently to reconstruct the 2015 Ukraine Power Grid cyberattack, as the first documented attack on critical OT infrastructure to cause a large-scale blackout, using a topology rebuilt from publicly available CIRT reporting, which provides a well-documented ground truth.

The remainder of the paper is organized as follows. Section 2 describes LM's \vortex{} attack surface assessment capability and introduces the reinforcement learning framework underlying \crow. Section 3 formalizes the Aalto attack graph modeling methodology, including node and edge semantics and the probabilistic  framework for attack propagation. Section 4 describes the use case of the 2015 Ukraine Power Grid scenario and its reconstruction. Section 5 presents the combined \vortex/\crow{} analysis and Section 6 presents the attack graph analysis, including baseline impact estimates, vulnerability-neutralization experiments and mitigation prioritization. Section 7 summarizes the key findings of this study and presents a plan for mitigation modeling.

\section{\textsc{Attack Surface Assessment With \vortex/\crow}}
\label{sec:vortexcrow}

\begin{figure}[htb]
    \centering
    \includegraphics[width=1\linewidth]{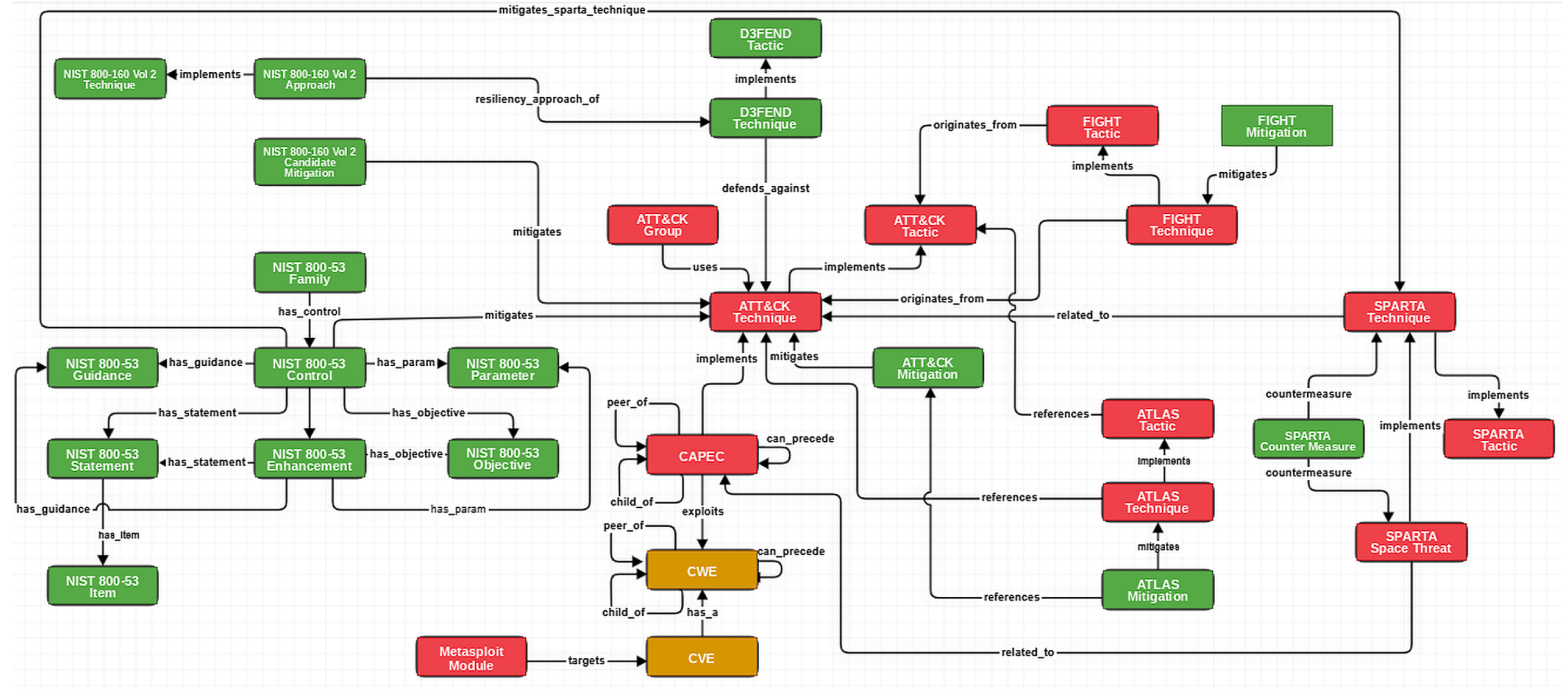}
    \caption{\vortex{} Cyber Knowledge Graph.}
    \label{fig:vortex_graph}
\end{figure}

In cyber incident attack vectors span every layer of a system, from embedded chipsets to network and account credentials, and extend to secondary paths through supporting systems. To address this complexity, LM developed \vortex, a scalable platform for software attack surface analysis, threat modeling, and mitigation reporting. \vortex{} builds a customized system knowledge graph (Fig.~\ref{fig:vortex_graph}) by ingesting system information, i.e. network configuration, Model-Based System Engineering (MBSE) data, Software Bills of Materials (SBOMs), and outputs from scanning tools such as ACAS, Coverity, and Fortify, and mapping them to CVE vulnerabilities, CWE weaknesses, CAPEC attack patterns, and mitigations drawn from recognized sources and frameworks such as NIST SP 800-53B and 800-160~\cite{nist800160}.
The knowledge base is updated periodically and is extensible to new sources and formats like MITRE ATT\&CK/D3FEND/ATLAS/FiGHT, and SPARTA~\cite{sparta}.

\begin{figure}
    \centering
    \includegraphics[width=1\linewidth]{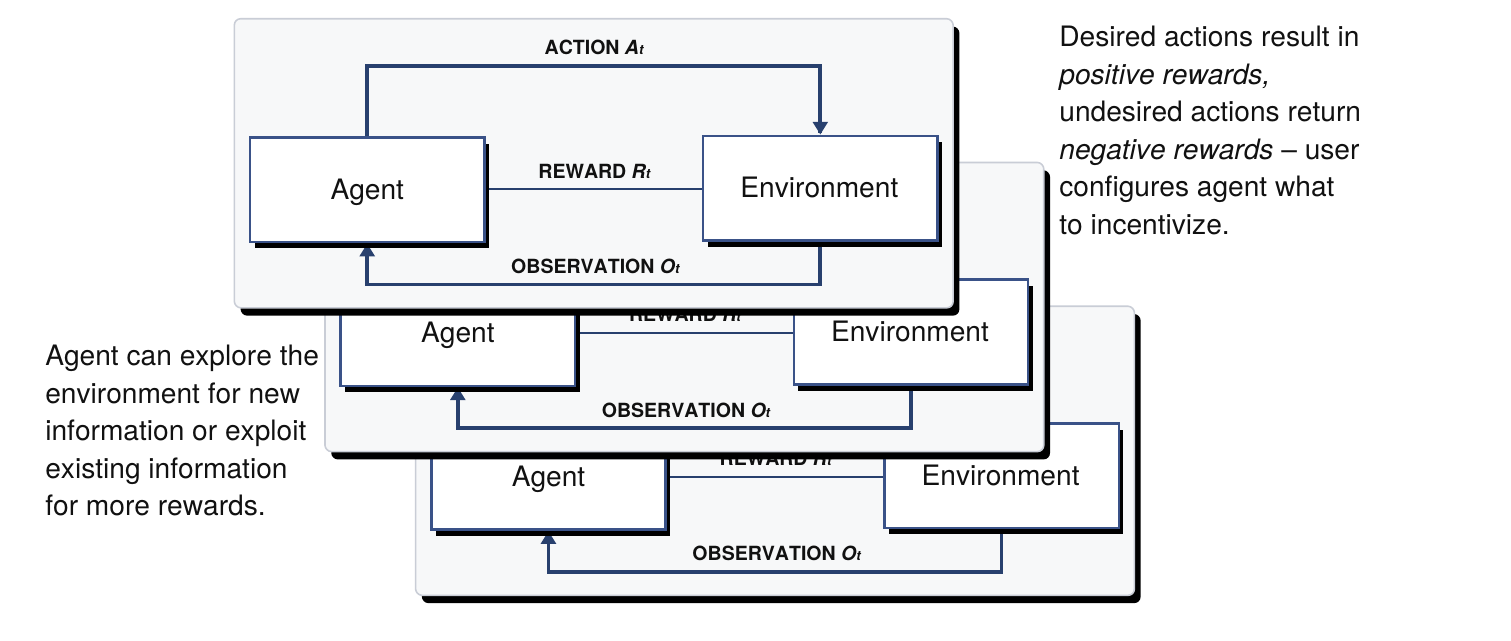}
    \caption{\acrfull{marl} Framework.}
    \label{fig:marl}
\end{figure}

Together CVE/CWE/CAPEC form the basis of the attack vectors that a reinforcement learning (RL) agent can apply against a given node in the network topology;
drawing cyber effects from the above mentioned authoritative resources is far more efficient and accurate than constructing vulnerabilities and malware from scratch. RL agents are trained in a synthetic environment guided by a reward-bearing objective function. Whereas human operators recognize only familiar attack patterns, LM's multi-agent reinforcement learning (MARL) framework (Fig.~\ref{fig:marl}) composes attack vectors and TTPs around specified cyber objectives, such as the Deny / Degrade / Disrupt / Deceive / Destroy (5D) effects or the confidentiality-integrity-availability (CIA) triad.

Trained over many thousands of episodes within the \vortex{} knowledge graph, the \crow{} agents inherit full traceability and standards compliance, and explainability is built in: each action is evaluated against the objective function during training. The approach has been validated by subject-matter experts from the Air Force Academy, Naval Academy and West Point, and integrated with commercial and government tools including STK, MATLAB, EXata, and AFSIM. By automating the otherwise labor-intensive cycle of discovering vulnerabilities, mapping them to exploits and testing attack vectors, \crow{} accelerates the assessment by orders of magnitude while covering the combinatorial attack space more comprehensively than manual analysis.

\section{Modeling Cybersecurity Defense with Attack Graphs}
\label{sec:attackgraph}

In cyber incidents, attackers typically compromise a network through an attack path, an ordered sequence of exploits, in which each step satisfies the prerequisites of the next, allowing lateral movement~\cite{SMILIOTOPOULOS2024e26317} from an entry point deeper into hosts and their applications. The collection of all feasible paths, constrained by reachability and system preconditions, forms an attack graph~\cite{lallie2020review,herttuainen2026}: a system-level representation that supports estimating an attacker's likelihood of success and evaluating mitigation strategies under resource constraints.

Existing work on attack graph analysis has mainly focused on either network-level or path-based risk assessment.
Probabilistic methods derive security scores by aggregating CVSS-based vulnerability measures across the network~\cite{wang2008attack,homer2013aggregating}, while graph-based approaches such as~\cite{stergiopoulos2022automatic} use closeness centrality metric and dependency paths to identify worst-case attack paths.
Metric suites such as~\cite{noel2017} further enable comparisons across enterprises via categorised family scores. While these approaches have advanced network vulnerability analysis, they do not account for per-service or per-host impact, which can be critical information for risk mitigation analysis.

Following the methodology of~\cite{herttuainen2026}, where new attack graph analysis metrics \ac{cei}, \ac{cnode}, \ac{cout}, and \ac{cin} were introduced, we model each exploit transition as an edge carrying a probabilistic exploitability weight, derived from vulnerability metrics, authentication structure, and observed attacker capabilities, which quantify the likelihood of a successful transition~\cite{wang2008attack}.
In this study we apply these four metrics with the following distinctions. First, the attack states that can be reached by exploiting vulnerabilities are represented as software instances present in the network devices (nodes). That is, we assume that exploiting any vulnerability in a software instance would lead to the same state during the attack.
While this is a clear simplification of reality, we argue that for a high-level analysis of where to focus one's defensive efforts, this approach is justified, particularly as the number of possible combinations becomes infeasible to compute when there are tens, if not hundreds of vulnerabilities per software instance. Since exploiting any vulnerability leads to the same state, the directed edge weight representing the likelihood of exploitation between software instances in the attack graph can be calculated as:
\begin{equation} \label{eq:prob_e}
    p_E(i,j) = 1 - \prod_{n \in V_j} \left[1 - \hat{e}(n)\right] \quad \Big| \quad \hat{e}(n) = \frac{e(n)}{E_{\max}}
\end{equation}
where $V_j$ is the set of all vulnerabilities present in software instance $j$, $E_{\max}$ is the maximum exploitability value in the scoring system, and $\hat{e}(n)\in[0,1]$ is the normalized exploitability score of vulnerability $n$.

Secondly, we assume full network-level connectivity between software instances that have a direct connection in the attack graph. This assumption is made because we do not have realistic data to estimate the weights of the network link ($w_n$). With this assumption, the probability of traversing a single attack step (originally a product of the exploitability probability $p_E$ and the network-level probability $p_N$) simply becomes
\begin{equation} \label{eq:prob_ovr}
        p(i,j) = p_{N}(i,j) * p_{E}(i,j) = p_{E}(i,j). \quad \Big| \quad w_n = 1 \implies p_N(i,j) = 1
\end{equation}

The last distinction from the original methodology is how the impact values are derived to compute the expected impact for the cumulative metrics. As all  vulnerabilities in a single software instance are aggregated, the impact values need to be aggregated as well.
To explore the full range of possible impacts, we have aggregated the impact scores taking the minimum, maximum, and mean across all vulnerabilities in a software instance. The expected impact is then defined as:
\begin{equation}
    E[I_{i}^{k}] = I^{k}(i) \sum_{j \in A} P(j,i), \quad k \in \{\min, \text{mean}, \max\}
\end{equation}
where $A$ is the set of all states immediately preceding the state $i$.
These aggregations are used in all cumulative metrics \ac{cei}, \ac{cnode}, \ac{cin}, and \ac{cout}.

Whereas the methodology in~\cite{herttuainen2026} only concentrated on acyclic graphs, in this study there is one cyclic pattern present in the attack graph between the nodes \emph{db\_server\_1} and \emph{engineering\_ws\_1}. When calculating the cumulative metrics, we have taken a self-avoiding path approach.
That is, only the first visit to a software instance is taken into account.
Similarly, with \ac{cin} and \ac{cout}, only the first time the attack enters or exits a node is taken into account.

Following the approach in~\cite{herttuainen2026}, we utilized the exploitability and impact values of \ac{cvss} version 2.0~\cite{cvss2} to estimate the probability and impact of the attack in its various states.
Although \ac{cvss}v2 is nowadays superseded by newer versions, in 2015 it was the de facto standard, as its successor \ac{cvss}v3.0, released in June 2015, had not yet been widely adopted at the time of the attack.
Moreover, a significant number of pre-2016 vulnerabilities in our scenario lack the \ac{cvss}v3.0 score.

IronMiner is built upon a cyber data lake with a wealth of publicly available information about CVEs including threat intelligence reports, the number of exploits available for each CVE, indicators that a CVE is under active exploitation and more. The LM IronMiner team used machine learning and statistical techniques to build a classification model that predicts the riskiest CVEs based on USG and industry reports of active exploitation. Each CVE in the NVD is assigned an IronMiner Threat Score between 1.0-10.0 and will be updated daily. IronMiner identifies and assigns high threat scores to CVEs in CISA's Known Exploited Vulnerability (KEV) catalog and those that are actively exploited or have other risky attributes detected with the machine learning. This methodology is found to correctly predict 87.3\% high threat CVEs in the KEV catalog and other risky vulnerabilities that threat actors have already exploited. IronMiner's Threat Score is based on researched and repeatable processes that inform defenders of CVE exploitation risks to improve the overall quality of CVE assessment. In addition, IronMiner automatically collects and processes new data daily to continuously monitor changes in CVE risk profiles; as such, it provides a higher amount of due diligence, consistency, and precision than manual CVE risk scoring systems. However, as IronMiner does not estimate the impact of vulnerabilities, \ac{cvss}v2 was used not only to provide a baseline for comparison but also to supply the impact estimates required by the attack graph analysis.

\section{Sample Scenario: 2015 Ukraine Power Grid}
\label{sec:scenario}

To test the combined methodology of \vortex /\crow{} and Attack Graph against a documented ground truth, we have applied them to the 2015 Ukraine Power Grid attack, reconstructing the attack vectors and performing the analysis from open-source incident reporting, principally the E-ISAC/CIRT report (\cite{eisac_sans_ukraine_2016}). On 23 December 2015, a roughly three-hour power outage struck the Ivano-Frankivsk region of Ukraine. Subsequently, malware was found in several substations (7 × 110 kV and 23 × 35 kV), and the outage, which affected up to 225,000 customers served by three regional distribution companies, was attributed to one or more Advanced Persistent Threat (APT) groups that leveraged BlackEnergy and related tools to disrupt the targeted substations simultaneously. The incident is significant as the first confirmed power blackout caused by a cyberattack. The major sequence of events list (MSEL) drawn from the CIRT report was as follows:
\begin{itemize}
    {\it{\small
    \item[1.]  Email phishing
    \item[2.] Reconnaissance and enumeration of the network to provide an initial backdoor
    \item[3.] Discovery and access Microsoft Active Directory® servers that contain corporate user accounts and credentials
    \item[4.] Use of encrypted tunnel from external network to get inside control system networks
    \item[5.] Discovery and access SCADA HMI due to an improperly configured firewall
    \item[6.] Control override of HMI operators and breaker opening command
    \item[7.] Several other actions with the intent to complicate the responses of control operators
    \item[8.] KillDisk malware attempting to wipe out the control center HMI and workstations
    }}
\end{itemize}

\begin{figure}[htb]
        \centering
        \includegraphics[width=0.9\linewidth]{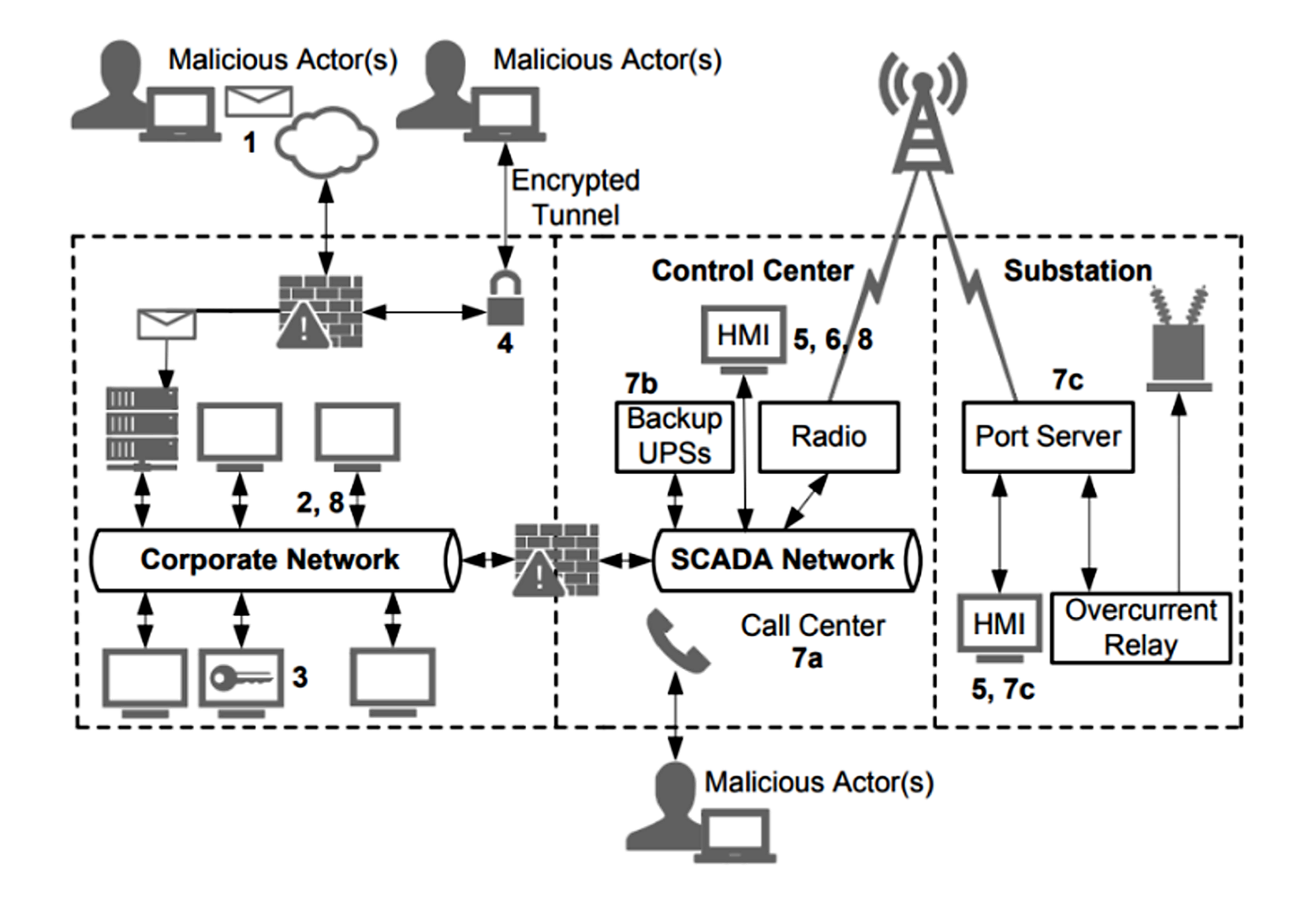}
        \caption{2015 Ukraine Power Grid Cyber Incident.}
        \label{fig:incident}
    \end{figure}

\begin{figure}[htb]
    \centering
    \includegraphics[width=1.0\textwidth,trim=1.6cm .7cm 2.2cm 0.2cm,clip]{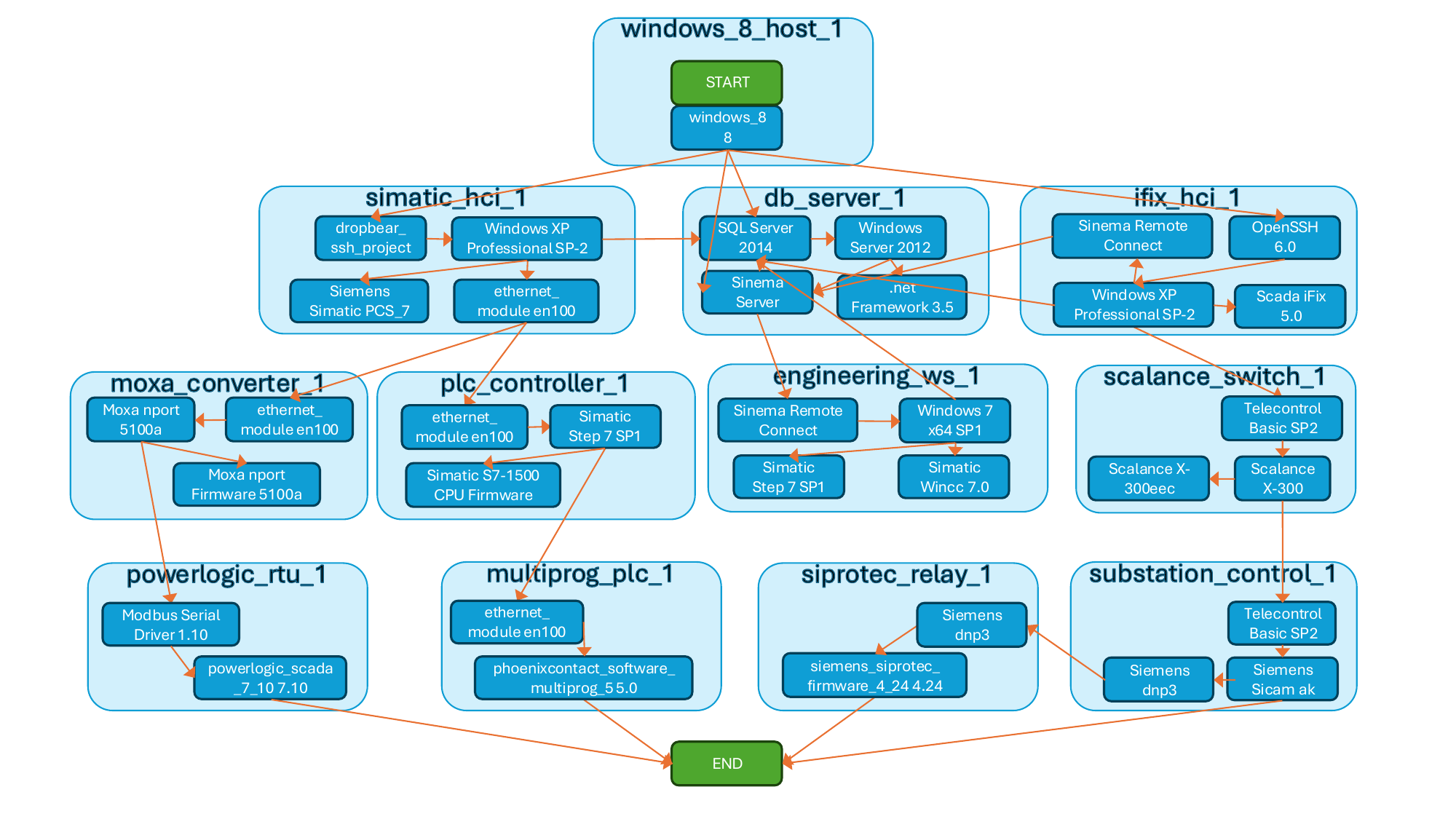}
    \caption{Assumed Power Grid and its ICS / OT Network Topology.}
    \label{fig:graph_topology}
\end{figure}

This sequence of events is mapped onto the representative grid architecture in Fig.~\ref{fig:incident}. Because the configuration of the Prykarpattya Oblenergo OT network was not disclosed in the CIRT report, we reconstructed a simplified topology under the stated hardware and software assumptions (Fig.~\ref{fig:graph_topology}). The left side of Fig.~\ref{fig:graph_topology} reflects period-appropriate 2015 components — e.g., programmable logic controllers (PLCs) and host controller interfaces (HCIs) — while the right side reflects a more recent configuration, allowing evaluation of whether multiple \crow{} agents select attack vectors comparable to those documented in the real incident under both legacy and modern assumptions.

\section{\vortex/\crow{} Analysis}
\label{sec:lm_analysis}

\begin{figure}
    \centering
    \includegraphics[width=\linewidth]{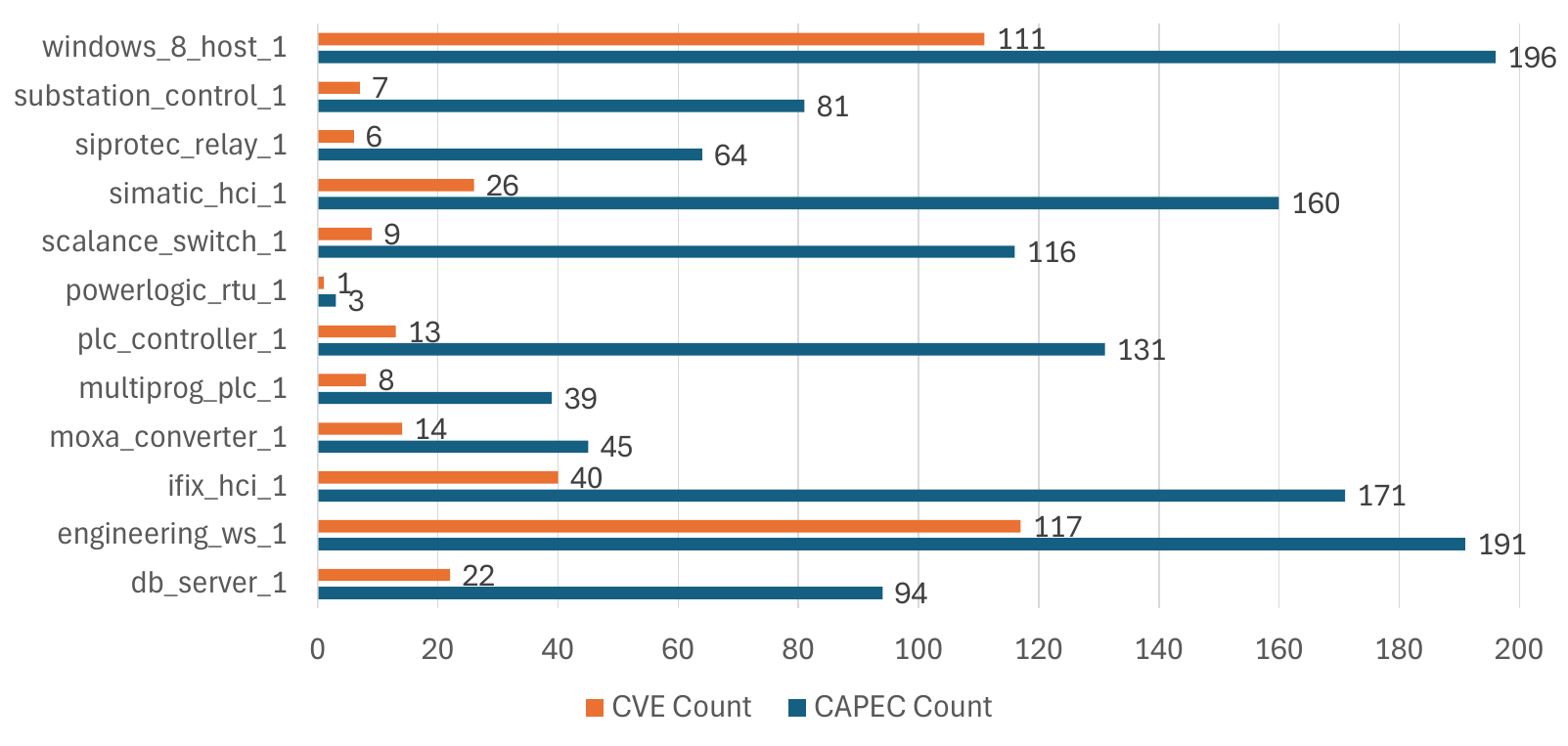}
    \caption{Vulnerability Assessment Results from \vortex.}
    \label{fig:vortex_res}
\end{figure}

Using the assumed hardware and software inventory of the reconstructed Ukraine Power Grid topology in Fig.~\ref{fig:graph_topology}, \vortex{} performed a vulnerability assessment that listed the candidate CVEs and CAPECs for each of its components, Fig.~\ref{fig:vortex_res}.
These define the action and observation spaces that the \crow{} agents explore to identify exploitable paths. CVE exploitability was scored with LM's IronMiner, chosen for its objectivity and its incorporation of recent intelligence, though it could be substituted by any other scoring method.

Because the 2015 adversary sought to disrupt power generation, the objective chosen for the reinforcement-learning reward used the specific CAPEC impacts of Bypass Protection Mechanism, Gain Privileges, and Execute Unauthorized Commands in that order for this initial pivot into the OT network. Open-source reporting attributes the incident to APT44 (Sandworm). The specific TTPs of this group (G0034) from MITRE ATT\&CK were leveraged from \vortex, and set as the prerequisite for all actions taken by the \crow{} agents. An example of a corresponding set of CVEs, CWEs, and CAPECs, as well as attack patterns for the given system configuration and 5D cyber effects, is shown in this sample set:

\begin{itemize}
{\small
    \item[]{('G0034', 'T1082', 'CAPEC-313', 'CWE-200', 'CVE-2007-2768')}
    \item[]{('G0034', 'T1539', 'CAPEC-31', 'CWE-20', 'CVE-2013-2811')}
    \item[]{('G0034', 'T1033', 'CAPEC-577', 'CWE-200', 'CVE-2015-1602')}
    \item[]{('G0034', 'T1078', 'CAPEC-560', 'CWE-522', 'CVE-2021-40360')}
}
\end{itemize}

\begin{figure}
    \centering
    \includegraphics[width=1\linewidth]{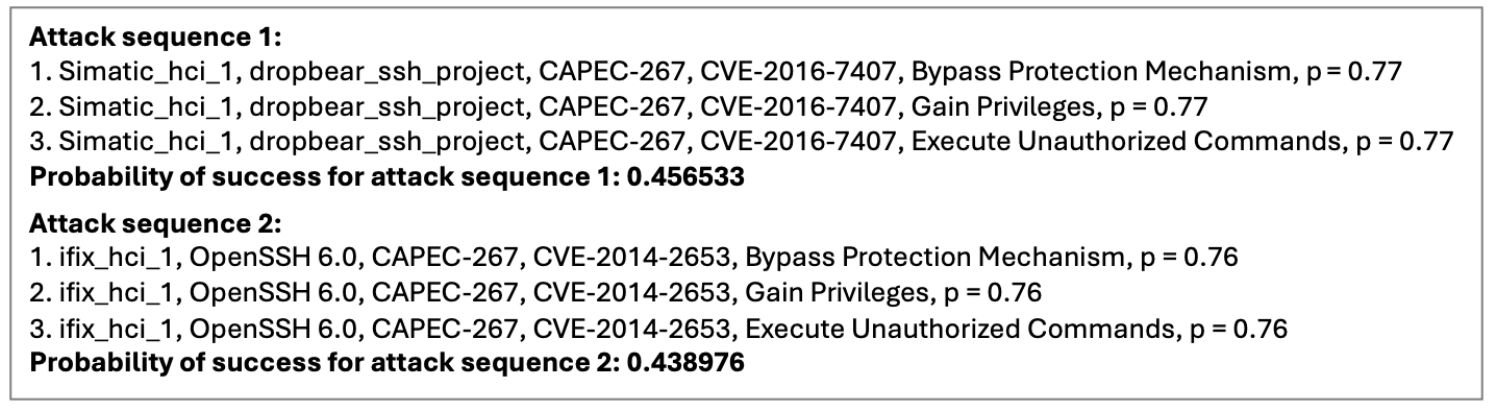}
    \caption{\crow{} results: two most probable or optimal attack sequences.}
    \label{fig:crow_res}
\end{figure}

Trained in the resulting action and observation spaces, \crow{} MARL agents learned a reward-maximizing policy that yields not only viable attack vectors but optimal attack sequences, emulating the behavior of APT44. The two highest-scoring sequences are shown in Fig.~\ref{fig:crow_res}. Each sequence's probability of success is the product of the per-step success probabilities of its constituent CVE–CAPEC pairs (scored with IronMiner), and the sequence with the highest overall probability is selected as the most likely course of action.
From the initial starting node of this scenario, the \crow{} agents prioritized attacking the node with a software configuration similar to the actual 2015 configuration, sequence 1, emulating the behavior of APT44 in the process.
An additional attack sequence 2 learned during training provides an alternative path that APT44 could have taken against a similar device with different software configurations.


\section{Attack Graph Analysis}
\label{sec:ag_analysis}

In Table~\ref{tab:cnode} we show the nodewise impact values using CVSSv2 and IronMiner exploitability values.
The \ac{cnode} results are consistent across CVSS and IronMiner: in both cases, the nodes~\emph{simatic\_hci\_1}, \emph{db\_server\_1}, and \emph{ifix\_hci\_1} have the highest potential impact. This is expected and explainable by the vicinity of the nodes to the entry point of the attack: the further the step is from the starting point of the attack, the lower its probability and, consequently, its expected impact.
Other explaining factors are the high number of vulnerabilities present in the software of these nodes, and the high number of attack steps within node \emph{db\_server\_1} as can be seen in Fig.~\ref{fig:graph_topology}.

\begin{table}[ht]
\centering
\caption{Vulnerability count and \ac{cnode} per device.}
\label{tab:cnode}
\begin{tabular}{|l|c|ccc|ccc|}
\hhline{--|------}
\multirow{3}{*}{Device} & \multirow{3}{*}{CVEs} & \multicolumn{6}{c|}{$C_{\text{NODE}}$} \\
\hhline{~~|---|---|}
 & & \multicolumn{3}{c|}{CVSS} & \multicolumn{3}{c|}{IronMiner} \\
\hhline{~~|---|---|}
 & & $I^{\min}$ & $I^{\text{mean}}$ & $I^{\max}$ & $I^{\min}$ & $I^{\text{mean}}$ & $I^{\max}$ \\
\hhline{--|------}
db\_server\_1          &           34 & \textbf{28.4} & \textbf{47.2} & \textbf{67.9} & \textbf{28.0} & \textbf{46.9} & \textbf{68.3} \\
engineering\_ws\_1     & \textbf{170} & 3.9           &           9.5 & 13.8          &           8.5 & 20.9          &          30.3 \\
ifix\_hci\_1           &           83 & 15.2          &          23.2 & 32.3          &          16.0 & 25.6          &          35.6 \\
moxa\_converter\_1     &           18 & 2.0           &           3.0 & 3.6           &           2.2 & 3.5           &           4.1 \\
multiprog\_plc\_1      &           10 & 6.9           &           9.0 & 10.9          &           4.8 & 6.2           &           7.5 \\
plc\_controller\_1     &           25 & 6.6           &          14.2 & 22.4          &           5.0 & 11.0          &          17.2 \\
powerlogic\_rtu\_1     &            3 & 3.1           &           3.7 & 4.4           &           3.7 & 4.6           &           5.5 \\
scalance\_switch\_1    &           13 & 0.8           &           0.9 & 1.0           &           6.2 & 7.3           &           8.0 \\
simatic\_hci\_1        &           48 & 12.1          &          20.8 & 32.7          &          12.0 & 20.8          &          32.7 \\
siprotec\_relay\_1     &            9 & 0.0           &           0.0 & 0.0           &           1.3 & 1.3           &           1.5 \\
substation\_control\_1 &           10 & 0.0           &           0.0 & 0.0           &           1.9 & 2.0           &           2.2 \\
\hhline{--|------}
\end{tabular}
\end{table}

\begin{figure}[htb]
    \centering
    \includegraphics[width=\textwidth]{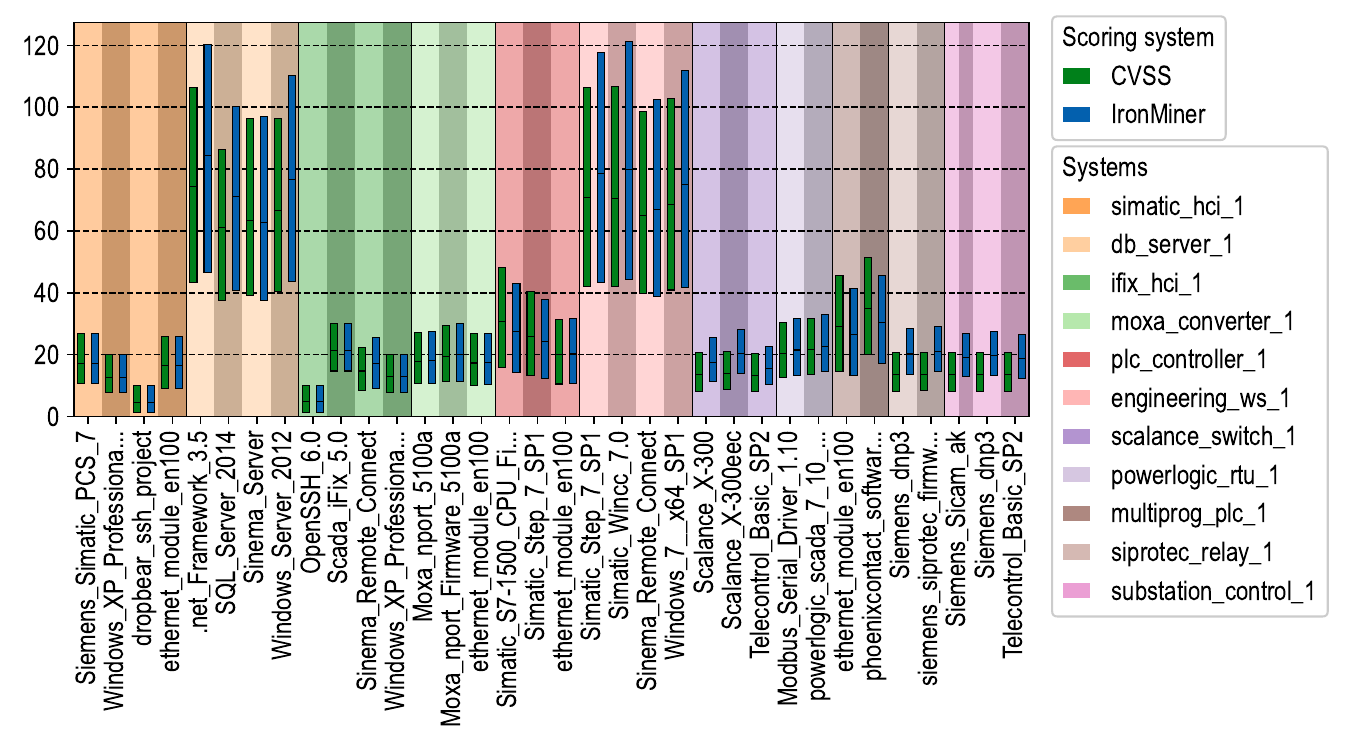}
    \caption{Expected cumulative impact (\ac{cei}) per software. The box plots show results obtained with minimum, mean, and maximum vulnerability impact values as the low end, midline, and high end, respectively.}
    \label{fig:fig16}
\end{figure}

The slight differences in the \ac{cnode} values between the two scoring systems are found on the right side of the attack graph: results obtained with IronMiner indicate higher impact on nodes \emph{scalance\_switch\_1}, \emph{siprotec\_relay\_1}, \emph{substation\_control\_1}, and \emph{engineering\_ws\_1} than those obtained with CVSS. This is clearly seen in the results represented in Fig.~\ref{fig:fig16}, in which the expected cumulative inbound impact per software is estimated using these two scoring systems.
Also, the cumulation of impact within a node can be witnessed in Fig.~\ref{fig:fig16} as, for example, the \ac{cei} of \emph{simatic\_hci\_1} increases when the attack propagates through its software Dropbear SSH, Windows XP, and en100 ethernet module.

\begin{figure}[h]
    \centering
    \subfloat[]{
        \includegraphics[width=0.5\textwidth,trim=0.2cm 0.2cm 0.2cm 0.2cm,clip]{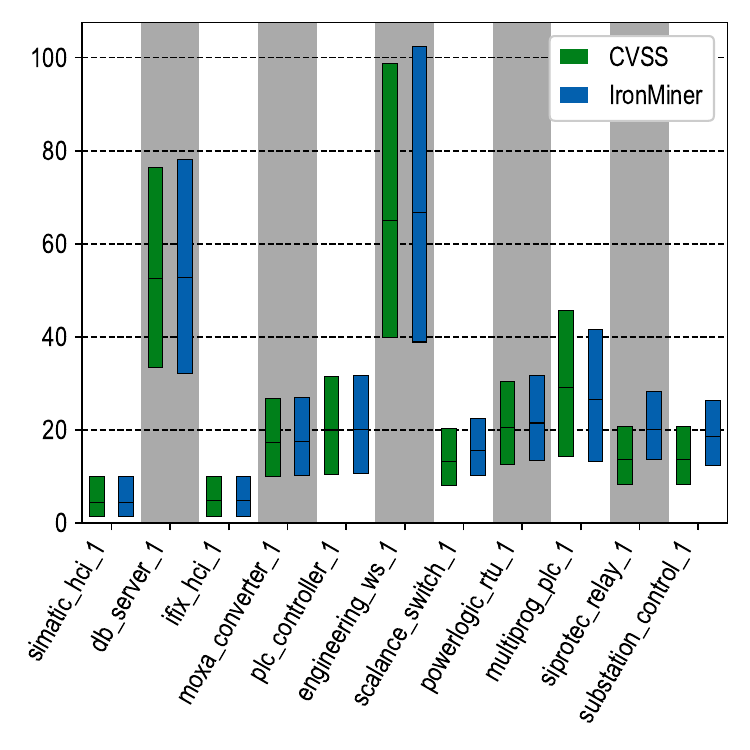}
        \label{fig:fig18}
    }
    \subfloat[]{
        \includegraphics[width=0.5\textwidth,trim=0.2cm 0.2cm 0.2cm 0.2cm,clip]{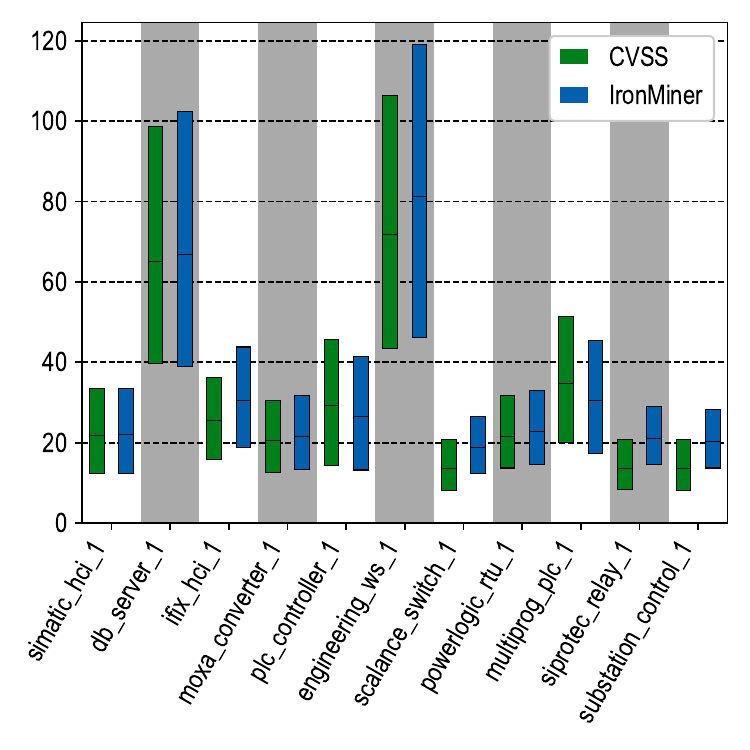}
        \label{fig:fig19}
    }
    \caption[]{\subref{fig:fig18} \ac{cin} and \subref{fig:fig19} \ac{cout} per node. The box plots show results obtained with minimum, mean, and maximum vulnerability impact values as the low end, midline, and high end, respectively.}
    \label{fig:baseline_cincout}
\end{figure}

The \ac{cin} and \ac{cout} results can be seen in Fig.~\ref{fig:baseline_cincout}.
In these box plots, the results obtained with $I^{\min}$ and $I^{\max}$ are represented by the low and high ends of the boxes, whereas those obtained with  $I^{\text{mean}}$ are represented by the middle lines of the boxes. These results are consistent with the \ac{cnode} results, and CVSS and IronMiner produce similar results. The only clear difference between these scoring systems is the slightly higher expected impacts on nodes \emph{scalance\_switch\_1}, \emph{siprotec\_relay\_1}, and \emph{substation\_control\_1}, on the right side of the attack graph and the node \emph{engineering\_ws\_1} when calculated using IronMiner.

\begin{figure}[htb]
    \centering
    \includegraphics[width=.6\textwidth]{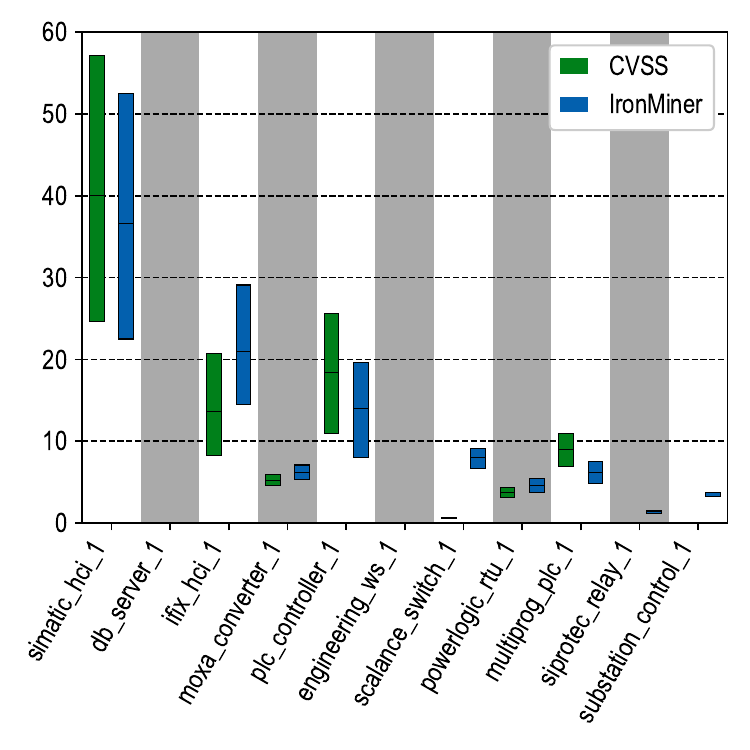}
    \caption{Reduction in expected cumulative impact of a successful attack (\ac{cei}[END]) after eliminating all vulnerabilities of a node. The box plots show results obtained with minimum, mean, and maximum vulnerability impact values as the low end, midline, and high end, respectively.}
    \label{fig:fig20}
\end{figure}

In order to find out the significance of each node in the network, we conducted a series of analyses in which the exploitability of all vulnerabilities associated with a given node was set to zero to mimic fully securing the node.
This process was repeated for each node individually, and the results of each analysis run were compared against the baseline to identify the most impactful nodes in the network.
In Fig.~\ref{fig:fig20} we show
the resulting box plot that illustrates
the decrease in overall attack impact ($CE[I_{END}]$) when the vulnerabilities of the individual nodes are neutralized. Those nodes that lead to a larger reduction in \ac{cei} can be interpreted as critical points in the attack graph, as they lie on multiple high-impact attack paths and may serve as pivotal points for attackers. Clearly, the node \emph{simatic\_hci\_1} on the left side of the attack graph is the most critical node for both CVSS and IronMiner scoring systems. The next most impactful nodes are \emph{ifix\_hci\_1} and \emph{plc\_controller\_1}, while securing the rest of the nodes results in moderate reductions. In particular, since \emph{db\_server\_1} and \emph{engineering\_ws\_1} are not on any of the attack paths that lead to the \emph{END} node, neutralizing their vulnerabilities has no effect on the overall impact of the attack. The most notable difference between the scoring systems is that IronMiner tends to have a higher reduction on the right side of the attack graph than CVSS.

Table~\ref{tab:node_cincout} shows the reduction in \ac{cin} and \ac{cout} after neutralizing the vulnerabilities of a node. In contrast to the \ac{cei} results, \emph{db\_server\_1} stands out for both measures due to the high number of attack paths and vulnerabilities, as established in the baseline. Compared to the results in Fig.~\ref{fig:baseline_cincout}, neutralizing \emph{engineering\_ws\_1} has little effect, implying that most of the high baseline \ac{cin} and \ac{cout} have cumulated before reaching the node. Unsurprisingly, the neutralization of the vulnerabilities of \emph{simatic\_hci\_1} and \emph{ifix\_hci\_1} has the most widespread effects, as they are the pivot points for the left and right sides of the attack graph, respectively. Securing either of these nodes reduces all \ac{cin} and \ac{cout} of the downstream nodes on the attack paths towards the \emph{END} node, while also partially reducing the impact on \emph{db\_server\_1} and \emph{engineering\_ws\_1}. Securing the nodes downstream of these pivotal points show diminishing results the further away they are from the start of the attack due to the impact already having cumulated before reaching these nodes. Moreover, consistent with the baseline and $CE[I_{END}]$ results, securing the peripheral nodes on the right side of the attack graph has no effect as measured by CVSS, while with IronMiner the reductions are minor but present.

\begin{table}[h]
\centering
\caption{Reduced \ac{cin} and \ac{cout} per node after securing a node}
\label{tab:node_cincout}
\textit{Node legend: }
A~=~simatic\_hci\_1,\;
B~=~db\_server\_1,\;
C~=~ifix\_hci\_1,\;
D~=~moxa\_converter\_1,\;
E~=~plc\_controller\_1,\;
F~=~engineering\_ws\_1,\;
G~=~scalance\_switch\_1,\;
H~=~powerlogic\_rtu\_1,\;
I~=~multiprog\_plc\_1,\;
J~=~siprotec\_relay\_1,\;
K~=~substation\_control\_1
\vspace{4pt}

\begin{tabular}{|l|l|ccc|ccc|ccc|ccc|}
\hhline{|-|-|---|---|---|---|}
\multirow{3}{*}{\makecell{Affected\\Node}} & \multirow{3}{*}{\makecell{Secured\\Node}} & \multicolumn{6}{c|}{$C_{\text{IN}}$} & \multicolumn{6}{c|}{$C_{\text{OUT}}$} \\
\hhline{~~|---|---|---|---|}
 & & \multicolumn{3}{c|}{CVSS} & \multicolumn{3}{c|}{IronMiner} & \multicolumn{3}{c|}{CVSS} & \multicolumn{3}{c|}{IronMiner} \\
\hhline{~~|---|---|---|---|}
 & & $I^{\min}$ & $I^{\text{mean}}$ & $I^{\max}$ & $I^{\min}$ & $I^{\text{mean}}$ & $I^{\max}$ & $I^{\min}$ & $I^{\text{mean}}$ & $I^{\max}$ & $I^{\min}$ & $I^{\text{mean}}$ & $I^{\max}$ \\
\hhline{|-|-|---|---|---|---|}
\multirow{4}{*}{A}
 & A &  1.4 &  4.5 & 10.0 &  1.4 &  4.5 & 10.0 & 12.4 & 21.9 & 33.4 & 12.3 & 22.0 & 33.5 \\
 & B &    - &    - &    - &    - &    - &    - &  1.0 &  1.0 &  1.0 &  0.7 &  0.7 &  0.7 \\
 & D &    - &    - &    - &    - &    - &    - &  1.0 &  1.0 &  1.0 &  1.0 &  1.0 &  1.0 \\
 & E &    - &    - &    - &    - &    - &    - &  1.3 &  3.6 &  5.7 &  1.4 &  3.8 &  5.9 \\
\hhline{|-|-|---|---|---|---|}
\multirow{4}{*}{B}
 & A &  8.8 & 13.7 & 21.0 &  8.5 & 13.4 & 20.7 &  8.8 & 13.7 & 21.0 &  8.6 & 13.6 & 21.0 \\
 & B & 17.3 & 25.2 & 34.1 & 15.2 & 22.8 & 32.6 & 23.6 & 37.7 & 56.4 & 21.9 & 37.0 & 56.9 \\
 & C & 15.6 & 25.2 & 35.9 & 16.3 & 28.1 & 41.3 & 15.6 & 25.2 & 35.9 & 16.4 & 28.4 & 41.8 \\
 & F &    - &    - &    - &    - &    - &    - &  0.5 &  1.7 &  2.3 &  1.4 &  4.3 &  5.8 \\
\hhline{|-|-|---|---|---|---|}
\multirow{3}{*}{C}
 & B &    - &    - &    - &    - &    - &    - &  7.3 & 10.5 & 13.5 &  7.2 & 10.9 & 15.7 \\
 & C &  1.4 &  4.8 & 10.0 &  1.4 &  4.8 & 10.0 & 15.9 & 25.5 & 36.1 & 18.8 & 30.6 & 43.8 \\
 & G &    - &    - &    - &    - &    - &    - &  0.3 &  0.3 &  0.3 &  2.5 &  2.5 &  2.5 \\
\hhline{|-|-|---|---|---|---|}
\multirow{3}{*}{D}
 & A & 10.1 & 17.3 & 26.7 & 10.2 & 17.5 & 26.9 & 12.6 & 20.5 & 30.5 & 13.4 & 21.5 & 31.8 \\
 & D &  1.0 &  1.0 &  1.0 &  1.0 &  1.0 &  1.0 &  3.4 &  4.1 &  4.8 &  4.2 &  5.0 &  5.9 \\
 & H &    - &    - &    - &    - &    - &    - &  2.0 &  2.6 &  3.3 &  2.6 &  3.4 &  4.3 \\
\hhline{|-|-|---|---|---|---|}
\multirow{3}{*}{E}
 & A & 10.5 & 20.0 & 31.4 & 10.6 & 20.2 & 31.8 & 14.4 & 29.2 & 45.7 & 13.2 & 26.5 & 41.5 \\
 & E &  1.3 &  3.6 &  5.7 &  1.4 &  3.8 &  5.9 &  5.2 & 12.8 & 19.9 &  4.0 & 10.0 & 15.6 \\
 & I &    - &    - &    - &    - &    - &    - &  1.2 &  3.3 &  5.2 &  0.9 &  2.3 &  3.6 \\
\hhline{|-|-|---|---|---|---|}
\multirow{4}{*}{F}
 & A &  8.8 & 13.7 & 21.0 &  8.6 & 13.6 & 21.0 &  8.8 & 13.7 & 21.0 &  8.6 & 13.6 & 21.0 \\
 & B & 23.6 & 37.7 & 56.4 & 21.9 & 37.0 & 56.9 & 27.2 & 44.4 & 64.1 & 29.2 & 51.3 & 73.4 \\
 & C & 15.6 & 25.2 & 35.9 & 16.4 & 28.4 & 41.8 & 15.6 & 25.3 & 35.8 & 16.6 & 28.7 & 42.1 \\
 & F &  0.5 &  1.7 &  2.3 &  1.4 &  4.3 &  5.8 &  4.1 &  8.4 & 10.0 &  8.6 & 18.6 & 22.3 \\
\hhline{|-|-|---|---|---|---|}
\multirow{3}{*}{G}
 & C &  8.1 & 13.3 & 20.3 & 10.3 & 15.5 & 22.5 &  8.2 & 13.6 & 20.7 & 12.4 & 18.7 & 26.4 \\
 & G &  0.3 &  0.3 &  0.3 &  2.5 &  2.5 &  2.5 &  0.4 &  0.6 &  0.7 &  4.6 &  5.7 &  6.4 \\
 & K &    - &    - &    - &    - &    - &    - &    - &    - &    - &  1.0 &  1.0 &  1.0 \\
\hhline{|-|-|---|---|---|---|}
\multirow{3}{*}{H}
 & A & 12.6 & 20.5 & 30.5 & 13.4 & 21.5 & 31.8 & 13.7 & 21.6 & 31.7 & 14.5 & 22.7 & 33.0 \\
 & D &  3.4 &  4.1 &  4.8 &  4.2 &  5.0 &  5.9 &  4.6 &  5.2 &  5.9 &  5.3 &  6.2 &  7.1 \\
 & H &  2.0 &  2.6 &  3.3 &  2.6 &  3.4 &  4.3 &  3.1 &  3.7 &  4.4 &  3.7 &  4.6 &  5.5 \\
\hhline{|-|-|---|---|---|---|}
\multirow{3}{*}{I}
 & A & 14.4 & 29.2 & 45.7 & 13.2 & 26.5 & 41.5 & 20.0 & 34.8 & 51.4 & 17.2 & 30.4 & 45.5 \\
 & E &  5.2 & 12.8 & 19.9 &  4.0 & 10.0 & 15.6 & 10.9 & 18.4 & 25.6 &  8.0 & 13.9 & 19.6 \\
 & I &  1.2 &  3.3 &  5.2 &  0.9 &  2.3 &  3.6 &  6.9 &  9.0 & 10.9 &  4.8 &  6.2 &  7.5 \\
\hhline{|-|-|---|---|---|---|}
\multirow{4}{*}{J}
 & C &  8.2 & 13.6 & 20.7 & 13.7 & 20.2 & 28.3 &  8.3 & 13.6 & 20.7 & 14.5 & 21.0 & 29.1 \\
 & G &  0.4 &  0.6 &  0.7 &  5.9 &  7.2 &  8.3 &  0.5 &  0.6 &  0.7 &  6.7 &  8.0 &  9.1 \\
 & J &    - &    - &    - &  0.4 &  0.5 &  0.7 &    - &    - &    - &  1.2 &  1.4 &  1.5 \\
 & K &    - &    - &    - &  2.3 &  2.5 &  2.9 &    - &    - &    - &  3.2 &  3.3 &  3.7 \\
\hhline{|-|-|---|---|---|---|}
\multirow{4}{*}{K}
 & C &  8.2 & 13.6 & 20.7 & 12.4 & 18.7 & 26.4 &  8.2 & 13.6 & 20.7 & 13.7 & 20.2 & 28.3 \\
 & G &  0.4 &  0.6 &  0.7 &  4.6 &  5.7 &  6.4 &  0.4 &  0.6 &  0.7 &  5.9 &  7.2 &  8.3 \\
 & J &    - &    - &    - &    - &    - &    - &    - &    - &    - &  0.4 &  0.5 &  0.7 \\
 & K &    - &    - &    - &  1.0 &  1.0 &  1.0 &    - &    - &    - &  2.3 &  2.5 &  2.9 \\
\hhline{|-|-|---|---|---|---|}
\end{tabular}
\end{table}

\section{Concluding Remarks}
\label{sec:summary}

Here, we have reported a study using LM's \vortex/\crow{} integrated knowledge graph \& multi-agent reinforcement learning framework and Aalto's probabilistic Attack Graph modeling approach to identify and prioritize attack vectors and their most likely sequences among all possible ones as well as to determine system-level risk metrics of vulnerabilities / exploits in OT networks. To compare these two approaches independently, we used the 2015 Ukraine Power Grid cyberattack as a well-documented realistic validation scenario.

One of the findings is that \vortex/\crow{} and Attack Graph frameworks produced similar results; both approaches converged on the same attack vectors and exploit sequences as documented in the incident record of the Ukraine Power Grid attack in 2015, thus providing mutual cross-validation.
This indicates that both holistic methodologies are useful in predicting future attacks in other scenarios or ICS / OT networks. Another observation that can be drawn from this study is that the CVSS and IronMiner scores produced similar results. IronMiner seems also to give actionable information. It should be noted that the version of CVSS used in this investigation was v2.0 which fits with the 2015 scenario timeframe, while IronMiner is based on a present-time assessment.

In the future, complicated challenges still exist; given the large number of potential attack vectors and attack sequences, how do we decide which vulnerabilities we can mitigate, given the constraints of time, cost and desired efficacy or reduction of risk? As a next step, the joint LM and Aalto team aims at proposing a holistic system-based Mitigation Prioritization Model to provide a solution. This will be done by developing a framework to optimize mitigation based on the desired efficacy, time, and cost constraints.

\bibliographystyle{unsrt}
\bibliography{references}

\begin{thebibliography}{10}

\bibitem{cve}
{MITRE}.
\newblock {Common Vulnerabilities and Exposures (CVE)}.
\newblock \url{https://cve.mitre.org}, 1999.
\newblock Accessed: 2026-06-01.

\bibitem{cwe}
{MITRE}.
\newblock {Common Weakness Enumeration (CWE)}.
\newblock \url{https://cwe.mitre.org}.
\newblock Accessed: 2026-06-01.

\bibitem{capec}
{MITRE}.
\newblock {Common Attack Pattern Enumeration and Classification (CAPEC)}.
\newblock \url{https://capec.mitre.org}.
\newblock Accessed: 2026-06-01.

\bibitem{nist80053}
{National Institute of Standards and Technology}.
\newblock {Security and Privacy Controls for Information Systems and Organizations}.
\newblock Technical Report SP 800-53B, NIST, 2020.

\bibitem{attck}
{MITRE}.
\newblock {MITRE ATT\&CK}.
\newblock \url{https://attack.mitre.org}.
\newblock Accessed: 2026-06-01.

\bibitem{d3fend}
{MITRE}.
\newblock {MITRE D3FEND}.
\newblock \url{https://d3fend.mitre.org}.
\newblock Accessed: 2026-06-01.

\bibitem{atlas}
{MITRE}.
\newblock {MITRE ATLAS}.
\newblock \url{https://atlas.mitre.org}.
\newblock Accessed: 2026-06-01.

\bibitem{fight}
{MITRE}.
\newblock {MITRE FiGHT}.
\newblock \url{https://fight.mitre.org}.
\newblock Accessed: 2026-06-01.

\bibitem{nist800160}
{National Institute of Standards and Technology}.
\newblock {Engineering Trustworthy Secure Systems}.
\newblock Technical Report SP 800-160 Vol. 2 Rev. 1, NIST, 2021.

\bibitem{sparta}
{The Aerospace Corporation}.
\newblock {Space Attack Research and Tactic Analysis (SPARTA)}.
\newblock \url{https://sparta.aerospace.org}.
\newblock Accessed: 2026-06-01.

\bibitem{SMILIOTOPOULOS2024e26317}
Christos Smiliotopoulos, Georgios Kambourakis, and Constantinos Kolias.
\newblock Detecting lateral movement: A systematic survey.
\newblock {\em Heliyon}, 10(4):e26317, 2024.

\bibitem{lallie2020review}
Harjinder~Singh Lallie, Kurt Debattista, and Jay Bal.
\newblock A review of attack graph and attack tree visual syntax in cyber security.
\newblock {\em Computer Science Review}, 35:100219, 2020.

\bibitem{herttuainen2026}
Joni Herttuainen, Vesa Kuikka, and Kimmo~K. Kaski.
\newblock Integrating network and attack graphs for service-centric impact analysis.
\newblock {\em IEEE Access}, pages 1--1, 2026.

\bibitem{wang2008attack}
Marcel Frigault, Lingyu Wang, Sushil Jajodia, and Anoop Singhal.
\newblock {\em Measuring the Overall Network Security by Combining CVSS Scores Based on Attack Graphs and Bayesian Networks}, pages 1--23.
\newblock Springer International Publishing, Cham, 2017.

\bibitem{homer2013aggregating}
John Homer, Su~Zhang, Xinming Ou, David Schmidt, Yanhui Du, S~Raj Rajagopalan, and Anoop Singhal.
\newblock Aggregating vulnerability metrics in enterprise networks using attack graphs.
\newblock {\em Journal of Computer Security}, 21(4):561--597, 2013.

\bibitem{stergiopoulos2022automatic}
George Stergiopoulos, Panagiotis Dedousis, and Dimitris Gritzalis.
\newblock Automatic analysis of attack graphs for risk mitigation and prioritization on large-scale and complex networks in industry 4.0.
\newblock {\em International Journal of Information Security}, 21(1):37--59, 2022.

\bibitem{noel2017}
Steven Noel and Sushil Jajodia.
\newblock {\em A Suite of Metrics for Network Attack Graph Analytics}, pages 141--176.
\newblock Springer International Publishing, Cham, 2017.

\bibitem{cvss2}
Peter Mell, Karen Scarfone, and Sasha Romanosky.
\newblock A complete guide to the common vulnerability scoring system version 2.0.
\newblock \url{https://www.first.org/cvss/v2/guide}, 2007.
\newblock Accessed: 2026-06-01.

\bibitem{eisac_sans_ukraine_2016}
{E-ISAC} and {SANS ICS}.
\newblock Analysis of the cyber attack on the {Ukrainian} power grid: Defense use case.
\newblock Technical report, Electricity Information Sharing and Analysis Center (E-ISAC) and SANS Industrial Control Systems, March 2016.

\end{thebibliography}

\end{document}